\documentclass[conference]{IEEEtran}
\usepackage{amsmath,amssymb,amsfonts}
\usepackage{algorithmic}
\usepackage{graphicx}
\usepackage{tikz}
\usepackage{pgfplots}
\usepackage{url}
\usepackage{xurl}
\usepackage{etoolbox}
\usepackage{xcolor}
\usepackage{hyperref}
\pgfplotsset{compat=1.18}

\appto\UrlBreaks{\do\/\do-}
\hypersetup{colorlinks=true, linkcolor=blue, citecolor=blue, filecolor=blue, urlcolor=blue}
\usetikzlibrary{positioning, arrows.meta, shapes, fit}
\title{Secure and Low-Latency IoT Analytics Using an Edge-Based Streaming Architecture}
\author{

\IEEEauthorblockN{Atul}
\IEEEauthorblockA{
Allenhouse Institute of Technology\\
Kanpur, India\\
Email: atulverma15704@gmail.com
}
\and
\IEEEauthorblockN{Varun Shukla}
\IEEEauthorblockA{
Allenhouse Institute of Technology\\
Kanpur, India\\
Email: varun.shuklaa@gmail.com
}

\and

\IEEEauthorblockN{Vivek Shukla}
\IEEEauthorblockA{
Allenhouse Institute of Technology\\
Kanpur, India\\
Email: vivekshukla0552@gmail.com
}
\and[\hfill\break\mbox{}\hfill]
\IEEEauthorblockN{Mehul Kumar Das}
\IEEEauthorblockA{
Allenhouse Institute of Technology\\
Kanpur, India\\
Email: dasmehulkumar08@gmail.com
}

}
\begin{document}
\maketitle

\begin{abstract}
The rapid growth of Internet of Things (IoT) devices has led to large-scale continuous data streams that require real-time processing. Traditional cloud-centric architectures fail to meet low-latency and bandwidth efficiency requirements due to network delays and high data transmission overhead. This paper proposes EdgeStream, a lightweight edge-based framework for real-time streaming analytics in IoT environments. The system integrates edge nodes for local processing with a cloud backend for coordination and storage, using MQTT-based communication and distributed processing with anomaly detection. Analytical models for latency, throughput, and bandwidth are developed to evaluate performance. Experimental results, compared to cloud-based systems, show up to 92.8
\end{abstract}


\section{Introduction}

The rapid expansion of the Internet of Things (IoT) has led to the deployment of billions of interconnected devices across domains such as smart cities, healthcare, industrial automation, and intelligent transportation systems. These devices continuously generate large volumes of streaming data that must be processed in real time to enable timely decision-making and automated responses~\cite{gubbi2013iot,zanella2014smart}.

Traditional IoT architectures primarily rely on centralized cloud computing for data storage and analytics. While the cloud offers high computational power and scalability, it introduces significant challenges for latency-sensitive applications. Data must travel from edge devices to remote data centers and back, resulting in increased response times, network congestion, and higher bandwidth consumption. Such limitations make cloud-centric approaches unsuitable for applications requiring immediate insights, such as autonomous systems, real-time health monitoring, and industrial control~\cite{shi2016edge,satyanarayanan2017emergence}.

Recent advancements in edge computing have emerged as a promising solution to address these limitations by bringing computation closer to the data source. By processing data locally at edge nodes, it is possible to reduce latency, minimize unnecessary data transmission, and improve system responsiveness. However, existing edge-based solutions often lack an integrated framework that effectively combines real-time streaming analytics, efficient communication protocols, and scalable system design. Additionally, many current approaches do not provide a comprehensive evaluation of performance trade-offs in terms of latency, throughput, and energy consumption~\cite{bonomi2012fog,chiang2016fog,mach2017mec,mao2017survey}.

To address these gaps, this paper proposes \textit{EdgeStream}, a lightweight and scalable edge-based framework designed for real-time IoT data analytics. The proposed system leverages a multi-layer architecture consisting of IoT devices, edge processing nodes, and a cloud backend. It integrates streaming analytics, lightweight anomaly detection, and efficient message communication using MQTT, enabling fast and reliable data processing at the edge while maintaining coordination with the cloud~\cite{banks2015mqtt,li2018learning}.

The performance of the proposed framework is evaluated through analytical modeling and experimental simulations, focusing on key metrics such as latency, throughput, and energy efficiency. The results indicate that the edge-based approach reduces latency and bandwidth usage while improving throughput in the evaluated IoT scenarios.

The main contributions of this paper are as follows:
\begin{itemize}
    \item We propose \textit{EdgeStream}, a novel edge-based framework for real-time streaming analytics in IoT environments.
    \item We design a multi-layer system architecture integrating IoT devices, edge nodes, and cloud infrastructure for efficient data processing.
    \item We develop mathematical models to analyze latency, throughput, and energy consumption in edge-enabled IoT systems.
    \item We implement a simulation-based testbed using real and virtual edge devices to evaluate system performance.
    \item We report latency reductions of up to 92.8\%, cloud-uplink bandwidth savings of 82--88\% per device-hour, and higher per-node throughput compared with cloud-only approaches.
\end{itemize}

\section{Related Work}

The challenges of latency, bandwidth consumption, and real-time processing in Internet of Things (IoT) systems have been widely studied, leading to the emergence of edge computing as a promising paradigm. Several research efforts have explored architectures, optimization techniques, and applications of edge-enabled IoT systems~\cite{shi2016edge,mach2017mec,mao2017survey}.

Early studies on IoT architectures primarily relied on cloud-centric models, where data generated by distributed devices is transmitted to centralized servers for processing and storage. While such approaches provide scalability and computational power, they suffer from significant latency and network congestion, making them unsuitable for time-sensitive applications. Traditional cloud-based IoT systems often experience delays due to long communication paths and heavy data traffic, limiting their effectiveness in real-time environments~\cite{gubbi2013iot,zanella2014smart}.

To address these limitations, edge computing has been introduced as a decentralized approach that processes data closer to its source. Edge computing reduces latency and bandwidth usage by enabling local data processing and minimizing unnecessary communication with the cloud. Edge-based architectures also improve system responsiveness, enhance privacy, and support real-time decision-making in applications such as smart cities, healthcare, and industrial automation~\cite{satyanarayanan2017emergence,bonomi2012fog,chiang2016fog,roman2018security}.

In addition to architectural advancements, research has focused on optimization techniques such as task offloading, resource allocation, and machine learning integration at the edge. Prior work has studied joint radio/computation optimisation, energy-aware offloading, balanced fog-cloud workload placement, and edge-assisted learning for IoT analytics~\cite{sardellitti2015joint,you2017energy,dinh2017offloading,deng2016optimal,li2018learning}. More generally, modern stream-processing systems have established design principles for low-latency analytics that remain highly relevant when such pipelines are adapted to resource-constrained edge environments~\cite{akidau2015dataflow,zaharia2013dstreams}.

Complementary research has also examined secure collaboration and trust mechanisms for distributed systems. In particular, key-sharing and authentication schemes based on non-commutative algebraic structures and polynomial constructions provide useful perspectives for protecting edge-to-cloud interactions, secure healthcare data exchange, and related IoT communication workflows~\cite{misra2019threeusers,shukla2021polynomials,chaturvedi2018threeparty,securePoly2023}. Recent studies have also discussed broader mathematical and logical foundations for AI-enabled computing~\cite{mathAI2024,mathLogic2024}, while adjacent analyses of decentralized digital ecosystems such as cryptocurrency highlight wider trust, governance, and policy considerations around distributed platforms~\cite{shukla2022cryptoindia}.

Despite these advancements, several limitations remain. First, many existing studies focus on either architectural design or performance optimization in isolation, lacking a unified framework that integrates real-time streaming analytics with efficient communication and processing mechanisms. Second, while latency reduction is widely reported, comprehensive evaluation across multiple performance metrics—such as throughput, bandwidth usage, and energy consumption—is often limited. Third, several works rely heavily on theoretical models or simulations without clearly defined experimental setups, making reproducibility and practical validation challenging.

To overcome these gaps, this paper proposes \textit{EdgeStream}, an integrated edge-based framework that combines real-time streaming analytics, lightweight anomaly detection, and efficient communication protocols within a unified architecture. Unlike existing approaches, the proposed work provides both analytical modeling and experimental evaluation across multiple performance metrics, offering a more comprehensive understanding of edge-enabled IoT systems.

\section{System Model and Problem Formulation}

This section formally defines the system architecture and formulates the problem addressed in this work.

\subsection{System Model}

We consider an edge-enabled IoT system composed of three layers: IoT devices, edge nodes, and a cloud backend. A large number of distributed IoT devices continuously generate streaming data that must be processed with minimal delay.

Let $\mathcal{D} = \{d_1, d_2, \ldots, d_n\}$ denote the set of IoT devices generating data streams. Each device produces a sequence of data packets over time. These data streams are transmitted to a set of edge nodes $\mathcal{E} = \{e_1, e_2, \ldots, e_m\}$ for local processing.

Each edge node performs:
\begin{itemize}
    \item Data filtering and aggregation
    \item Real-time analytics
    \item Lightweight anomaly detection
\end{itemize}

The cloud layer $\mathcal{C}$ is responsible for:
\begin{itemize}
    \item Long-term data storage
    \item Global coordination
    \item Computationally intensive analytics
\end{itemize}

\subsection{Architecture Overview}

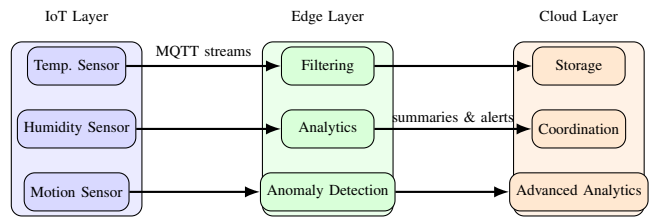
\begin{figure}[htbp]
    \centering
    \begin{tikzpicture}[
        scale=0.72,
        transform shape,
        node distance=1.4cm,
        every node/.style={font=\footnotesize},
        layer/.style={draw, rounded corners, minimum width=2.4cm, minimum height=3.2cm, align=center},
        box/.style={draw, rounded corners, minimum width=1.7cm, minimum height=0.7cm, align=center, fill=white},
        arrow/.style={-{Latex[length=2mm]}, thick}
    ]
        \node[layer, fill=blue!8] (iotlayer) {};
        \node[layer, fill=green!8, right=2.2cm of iotlayer] (edgelayer) {};
        \node[layer, fill=orange!10, right=2.2cm of edgelayer] (cloudlayer) {};

        \node[above=0.2cm of iotlayer.north] {IoT Layer};
        \node[above=0.2cm of edgelayer.north] {Edge Layer};
        \node[above=0.2cm of cloudlayer.north] {Cloud Layer};

        \node[box, fill=blue!15, above=0.8cm of iotlayer.center] (sensor1) {Temp. Sensor};
        \node[box, fill=blue!15, at=(iotlayer.center)] (sensor2) {Humidity Sensor};
        \node[box, fill=blue!15, below=0.8cm of iotlayer.center] (sensor3) {Motion Sensor};

        \node[box, fill=green!15, above=0.8cm of edgelayer.center] (edge1) {Filtering};
        \node[box, fill=green!15, at=(edgelayer.center)] (edge2) {Analytics};
        \node[box, fill=green!15, below=0.8cm of edgelayer.center] (edge3) {Anomaly Detection};

        \node[box, fill=orange!18, above=0.8cm of cloudlayer.center] (cloud1) {Storage};
        \node[box, fill=orange!18, at=(cloudlayer.center)] (cloud2) {Coordination};
        \node[box, fill=orange!18, below=0.8cm of cloudlayer.center] (cloud3) {Advanced Analytics};

        \draw[arrow] (sensor1.east) -- node[above]{MQTT streams} (edge1.west);
        \draw[arrow] (sensor2.east) -- (edge2.west);
        \draw[arrow] (sensor3.east) -- (edge3.west);
        \draw[arrow] (edge2.east) -- node[above]{summaries \& alerts} (cloud2.west);
        \draw[arrow] (edge1.east) -- (cloud1.west);
        \draw[arrow] (edge3.east) -- (cloud3.west);
    \end{tikzpicture}
    \caption{Proposed EdgeStream architecture for real-time IoT analytics.}
    \label{fig:architecture}
\end{figure}

Figure~\ref{fig:architecture} illustrates the overall system architecture. IoT devices transmit data to nearby edge nodes using the MQTT lightweight communication protocol~\cite{banks2015mqtt}. Edge nodes process data locally and forward only relevant information to the cloud.

\subsection{Problem Statement}

Given a set of IoT devices generating continuous data streams, the objective is to design a system that minimizes end-to-end latency and bandwidth usage while maximizing throughput and energy efficiency.

Formally, the problem can be expressed as:

\begin{equation}
\min \; L, \quad \min \; B, \quad \max \; T, \quad \min \; E
\end{equation}

where:
\begin{itemize}
    \item $L$ = End-to-end latency
    \item $B$ = Bandwidth consumption
    \item $T$ = Throughput
    \item $E$ = Energy consumption
\end{itemize}

These objectives are interdependent and require a trade-off between local processing at the edge and centralized processing in the cloud.

For scalar comparison of candidate operating points, the raw objectives are later converted into dimensionless normalized scores before they are combined into a single utility value.

\subsection{Assumptions}

The system model is based on the following assumptions:

\begin{itemize}
    \item IoT devices generate continuous data streams at a fixed or variable rate.
    \item Edge nodes have limited computational and storage resources compared to the cloud.
    \item Network communication between IoT devices and edge nodes has lower, but non-zero, latency than communication with the remote cloud.
    \item MQTT is used as a lightweight and reliable communication protocol.
    \item Data preprocessing at the edge reduces the volume of data transmitted to the cloud.
    \item Time-critical acknowledgements or control actions can be generated at the edge, while cloud uploads of summaries occur asynchronously unless explicit cloud coordination is required.
\end{itemize}

\subsection{Constraints}

The design of the system is subject to the following constraints:

\begin{itemize}
    \item \textbf{Resource Constraints:} Edge devices have limited CPU, memory, and power.
    \item \textbf{Latency Requirements:} Applications require response times in the order of milliseconds.
    \item \textbf{Bandwidth Limitations:} Continuous transmission of raw data to the cloud is not feasible.
    \item \textbf{Scalability:} The system must support a large number of IoT devices.
    \item \textbf{Energy Efficiency:} Devices must operate under limited power budgets.
\end{itemize}

\subsection{Design Objective}

Based on the above formulation, the goal of this work is to develop an edge-based framework that performs efficient local processing to reduce latency and bandwidth consumption while maintaining high throughput and energy efficiency.

\section{Proposed Method: EdgeStream Framework}

This section presents \textit{EdgeStream}, a lightweight and scalable edge-based framework designed for real-time IoT data analytics. The primary objective of EdgeStream is to minimize latency and bandwidth usage while enabling efficient, real-time decision-making at the edge.

\subsection{Framework Overview}

EdgeStream follows a multi-layer architecture consisting of IoT devices, edge nodes, and a cloud backend. Unlike traditional cloud-centric systems, the proposed framework performs data processing as close to the data source as possible, thereby reducing communication overhead and response time.

The framework integrates streaming analytics, lightweight anomaly detection, and efficient communication using a publish-subscribe model~\cite{akidau2015dataflow,zaharia2013dstreams,banks2015mqtt}.

\subsection{Architecture Components}

The key components of the EdgeStream framework are described below:

\begin{itemize}
    \item \textbf{IoT Layer:} Consists of distributed sensors and devices that continuously generate streaming data such as temperature, humidity, and motion readings.

    \item \textbf{Edge Processing Layer:} Edge nodes (e.g., Raspberry Pi and Jetson Nano) perform local data processing, including filtering, aggregation, and real-time analytics.

    \item \textbf{Streaming Engine:} Handles continuous data streams and supports real-time processing using lightweight mechanisms suitable for resource-constrained environments~\cite{akidau2015dataflow,zaharia2013dstreams}.

    \item \textbf{Communication Module:} Implements MQTT-based publish-subscribe messaging to ensure low-latency and efficient data transfer between system components~\cite{banks2015mqtt}.

    \item \textbf{Cloud Layer:} Responsible for long-term storage, system coordination, and execution of computationally intensive tasks.
\end{itemize}

\subsection{Data Flow Mechanism}

The data flow in EdgeStream is designed to reduce unnecessary data transmission and enable fast local decision-making:

\begin{enumerate}
    \item IoT devices generate continuous data streams.
    \item Data is transmitted to the nearest edge node via MQTT.
    \item Edge nodes preprocess data by removing noise and aggregating relevant information.
    \item A lightweight anomaly detection module analyzes the processed data in real time.
    \item If an anomaly or critical event is detected, immediate action is triggered locally.
    \item Only summarized or relevant data is forwarded to the cloud for storage and further analysis.
\end{enumerate}

\subsection{Lightweight Anomaly Detection Algorithm}

To enable real-time analysis at the edge, a lightweight statistical anomaly detection method is employed. The approach is based on z-score normalization, which is computationally efficient and consistent with prior lightweight anomaly-detection practice in streaming and network-monitoring settings~\cite{chandola2009anomaly,ahmed2016network}.

The detector is applied independently to each device stream using a fixed-length sliding window of $W$ samples. To avoid conflict with the service-rate notation $\mu$ used later in the queueing model, the detector's window mean and standard deviation are denoted by $m_t$ and $s_t$, respectively.

Let $x_t$ denote the current sample from a given device and let $\mathcal{W}_{t-1} = \{x_{t-W}, x_{t-W+1}, \ldots, x_{t-1}\}$ denote the previous window of $W$ samples. The window mean and sample standard deviation are computed as:

\begin{equation}
m_{t-1} = \frac{1}{W} \sum_{i=t-W}^{t-1} x_i
\end{equation}

\begin{equation}
s_{t-1} = \sqrt{\frac{1}{W-1} \sum_{i=t-W}^{t-1} \left(x_i - m_{t-1}\right)^2 + \varepsilon}
\end{equation}

where $\varepsilon > 0$ is a small variance floor used to avoid division by zero. The z-score for the new sample is then computed against the previous window as:

\begin{equation}
z_t = \frac{x_t - m_{t-1}}{s_{t-1}}
\end{equation}

A data point is classified as anomalous if:

\begin{equation}
|z_t| > \theta
\end{equation}

where $\theta$ is a predefined threshold. After scoring $x_t$, the oldest sample in the window is discarded and $x_t$ is inserted. During initialization, the first $W$ samples of each device stream are used only to populate the window and are not flagged as anomalies.

\subsection{Algorithm Workflow}

The overall processing workflow of EdgeStream is summarized below:

\begin{algorithmic}[1]
\STATE Initialize threshold $\theta$, window length $W$, and a per-device sliding buffer
\WHILE{the data stream is active}
    \STATE Receive a new data sample $x_t$ from device $d_i$
    \STATE Apply filtering and aggregation
    \IF{the buffer for $d_i$ contains fewer than $W$ samples}
        \STATE Append $x_t$ to the buffer for $d_i$
    \ELSE
        \STATE Compute $m_{t-1}$ and $s_{t-1}$ from the previous $W$ buffered samples
        \STATE Compute $z_t = (x_t - m_{t-1}) / s_{t-1}$
        \IF{$|z_t| > \theta$}
            \STATE Trigger an alert or local action
        \ELSE
            \STATE Forward processed data to the cloud if required
        \ENDIF
        \STATE Append $x_t$ and discard the oldest buffered sample for $d_i$
    \ENDIF
\ENDWHILE
\end{algorithmic}

\subsection{Key Features of the Framework}

\begin{itemize}
    \item \textbf{Low Latency:} Processing at the edge reduces response time significantly.
    \item \textbf{Bandwidth Efficiency:} Only relevant data is transmitted to the cloud.
    \item \textbf{Scalability:} Distributed edge nodes support large-scale IoT deployments.
    \item \textbf{Energy Efficiency:} Reduced communication leads to lower energy consumption.
\end{itemize}

\section{Mathematical Model}

To evaluate the performance of the proposed \textit{EdgeStream} framework, we model four key metrics: latency, bandwidth, throughput, and energy consumption. These models provide a theoretical basis for comparing edge-based and cloud-based IoT systems.

\subsection{Latency Model}

Latency represents the total time between message generation at an IoT device and completion of the time-critical response associated with that message.

To avoid ambiguous notation, let $L_{\text{comm}}^{a \rightarrow b}$ denote the one-way communication delay from node $a$ to node $b$, including transmission, access, and propagation components. Let $L_{\text{q},x}$ and $L_{\text{p},x}$ denote queueing and processing delays at node $x$, where $x \in \{e,c\}$ corresponds to an edge node or the cloud.

In a traditional cloud-based system, latency can be expressed as:

\begin{equation}
L_{\text{cloud}} = L_{\text{comm}}^{d \rightarrow c} + L_{\text{q},c} + L_{\text{p},c} + L_{\text{comm}}^{c \rightarrow d}
\end{equation}

where:
\begin{itemize}
    \item $L_{\text{comm}}^{d \rightarrow c}$ = Communication delay from the device to the cloud
    \item $L_{\text{q},c}$ = Queueing delay at the cloud
    \item $L_{\text{p},c}$ = Processing delay at the cloud
    \item $L_{\text{comm}}^{c \rightarrow d}$ = Response-path communication delay from the cloud to the device
\end{itemize}

In the proposed edge-based system, the time-critical path still includes non-zero communication between the device and the edge node:

\begin{equation}
L_{\text{edge}} = L_{\text{comm}}^{d \rightarrow e} + L_{\text{q},e} + L_{\text{p},e} + L_{\text{comm}}^{e \rightarrow d}
\end{equation}

If a message additionally requires synchronous cloud coordination, the latency becomes:

\begin{equation}
L_{\text{edge+cloud}} = L_{\text{edge}} + L_{\text{comm}}^{e \rightarrow c} + L_{\text{q},c}^{\text{sync}} + L_{\text{p},c}^{\text{sync}} + L_{\text{comm}}^{c \rightarrow e}
\end{equation}

Since edge nodes are geographically and topologically closer to IoT devices, we expect

\begin{equation}
L_{\text{comm}}^{d \rightarrow e} + L_{\text{comm}}^{e \rightarrow d} < L_{\text{comm}}^{d \rightarrow c} + L_{\text{comm}}^{c \rightarrow d}
\end{equation}

rather than assuming the edge communication terms are zero. In the evaluated EdgeStream pipeline, local acknowledgement or actuation completes at the edge and cloud upload is asynchronous for the latency-critical path, so the measured end-to-end delay corresponds to $L_{\text{edge}}$.

\subsection{Throughput Model}

Throughput is defined as the number of data messages processed per unit time:

\begin{equation}
T = \frac{N}{\Delta t}
\end{equation}

where:
\begin{itemize}
    \item $T$ = Throughput (messages per second)
    \item $N$ = Number of processed messages
    \item $\Delta t$ = Time interval
\end{itemize}

To account for system load, we model queueing delay using an M/M/1 queue:

\begin{equation}
D_q = \frac{\lambda}{\mu(\mu - \lambda)}, \quad \lambda < \mu
\end{equation}

where:
\begin{itemize}
    \item $\lambda$ = Arrival rate
    \item $\mu$ = Service rate
\end{itemize}

For a stable M/M/1 system, the long-run throughput equals the departure rate, which matches the arrival rate when $\lambda < \mu$:

\begin{equation}
T_{\text{eff}} = \lambda = \rho \mu, \quad \rho = \frac{\lambda}{\mu}, \quad \lambda < \mu
\end{equation}

The quantity $\mu - \lambda$ represents spare service capacity rather than throughput. Therefore, the maximum sustainable throughput of a single processing node is upper-bounded by its service rate $\mu$, and for $k$ balanced parallel edge nodes the aggregate sustainable throughput satisfies $T_{\text{agg}} \leq k\mu$.

Edge computing improves throughput by increasing aggregate service capacity across multiple nodes and by keeping each node's utilization below saturation, which also reduces queueing delay.

\subsection{Bandwidth Model}

Cloud-uplink bandwidth is defined as the average number of bytes transmitted toward the cloud per unit time. Let $r$ denote the message generation rate, $s$ the mean serialized payload size, $h$ the average protocol overhead per message, and $\alpha \in [0,1]$ the forwarding ratio after edge-side filtering and aggregation. Then,

\begin{equation}
B = \alpha r (s + h)
\end{equation}

where:
\begin{itemize}
    \item $r$ = Message generation rate
    \item $s$ = Mean payload size
    \item $h$ = Per-message communication overhead
    \item $\alpha$ = Fraction of messages forwarded to the cloud
\end{itemize}

For the cloud-only baseline, every message is uploaded for remote processing, so

\begin{equation}
B_{\text{cloud}} = r (s + h)
\end{equation}

For EdgeStream, only a subset of filtered or summarized messages is uploaded, giving

\begin{equation}
B_{\text{edge}} = \alpha_{\text{edge}} r (s + h), \quad 0 \leq \alpha_{\text{edge}} < 1
\end{equation}

If $n$ devices with similar traffic characteristics are active, the aggregate cloud-uplink bandwidth is $B_{\text{agg}} = nB$. The per-device-hour quantity reported in Section~\ref{sec:results} is obtained by integrating $B$ over one hour and normalizing by the number of active devices.

\subsection{Energy Consumption Model}

Energy efficiency is critical in IoT systems with constrained resources. The total energy consumption per device is given by:

\begin{equation}
E_{\text{total}} = E_{\text{tx}} + E_{\text{rx}} + E_{\text{proc}}
\end{equation}

Each component can be expressed as:

\begin{equation}
E = P \times t
\end{equation}

where:
\begin{itemize}
    \item $P$ = Power consumption
    \item $t$ = Duration of operation
\end{itemize}

In edge-based systems, energy consumption is reduced because less raw data is transmitted to the cloud. Therefore, transmission and reception energy ($E_{\text{tx}}$, $E_{\text{rx}}$) are significantly lower compared to cloud-based systems.

\subsection{Trade-Off Analysis}

To capture the four-way trade-off between latency, bandwidth, throughput, and energy consumption, we define dimensionless performance scores relative to reference values measured under the same workload. Let $L_{\text{ref}}$, $B_{\text{ref}}$, $T_{\text{ref}}$, and $E_{\text{ref}}$ denote these reference values, which in our experiments correspond to the cloud-only baseline.

\begin{equation}
S_L = \frac{L_{\text{ref}}}{L}, \quad
S_B = \frac{B_{\text{ref}}}{B}, \quad
S_T = \frac{T}{T_{\text{ref}}}, \quad
S_E = \frac{E_{\text{ref}}}{E}
\end{equation}

Using these normalized scores, the scalar utility is defined as:

\begin{equation}
U = w_L S_L + w_B S_B + w_T S_T + w_E S_E
\end{equation}

where:
\begin{itemize}
    \item $S_L, S_B, S_T, S_E$ = Dimensionless scores for latency, bandwidth, throughput, and energy, respectively
    \item $w_L, w_B, w_T, w_E \geq 0$ = Application-dependent weights
    \item $w_L + w_B + w_T + w_E = 1$
\end{itemize}

This formulation preserves all four objectives from the problem statement while avoiding dimensional inconsistency, because each score is normalized before combination. Larger values of $U$ indicate better overall operating points for the chosen application priorities. These models are validated through the experimental results presented in the following sections.

\section{Implementation and Experimental Setup}

This section describes the implementation details and experimental configuration used to evaluate the proposed \textit{EdgeStream} framework. The setup is designed to ensure reproducibility and fair comparison with a cloud-centric baseline.

\subsection{Hardware Configuration}

The experimental testbed consists of heterogeneous edge devices and a cloud server:

\begin{itemize}
    \item \textbf{Edge Devices:}
    \begin{itemize}
        \item Raspberry Pi 4 Model B (4GB RAM, quad-core Cortex-A72 @ 1.5 GHz)
        \item NVIDIA Jetson Nano (4GB RAM, 128-core GPU)
        \item Intel NUC (16GB RAM, Intel i5 @ 2.7 GHz)
        \item Two additional virtual edge instances mirroring the Raspberry Pi software stack for the five-node scaling experiments
    \end{itemize}

    \item \textbf{Cloud Server:}
    \begin{itemize}
        \item Virtual machine with 8 vCPUs and 16GB RAM
    \end{itemize}
\end{itemize}

\subsection{Software Stack}

The system is implemented using the following software components:

\begin{itemize}
    \item \textbf{Programming Language:} Python 3.10
    \item \textbf{Messaging Protocol:} MQTT using Eclipse Mosquitto broker
    \item \textbf{Libraries:} NumPy (data processing), Matplotlib (visualization), and Paho-MQTT (communication)
    \item \textbf{Operating System:} Ubuntu 22.04 (cloud), Raspberry Pi OS / Linux (edge devices)
\end{itemize}

\subsection{Dataset and Workload Generation}

Since real-world IoT datasets with continuous streaming behavior are limited, a synthetic workload is generated to simulate realistic IoT scenarios.

\begin{itemize}
    \item Sensor types: temperature, humidity, and vibration (accelerometer)
    \item Data generation rate: 100--1000 messages per second per device
    \item Number of devices: 50--500 simulated IoT nodes
    \item Data format: JSON-based streaming messages
\end{itemize}

To emulate real-world variability, random noise and anomalies are injected into the data stream based on predefined distributions.

\subsection{System Parameters}

The following parameters are used in the experiments:

\begin{itemize}
    \item MQTT QoS level: 1
    \item Anomaly detection threshold ($\theta$): 3.0 (z-score based)
    \item Sliding-window size ($W$): 50 samples per device stream
    \item Variance floor ($\varepsilon$): $10^{-6}$
    \item Message size: 256 bytes to 1 KB
    \item Network latency (simulated cloud): 100--500\,ms
    \item Edge processing delay: 5--20\,ms
\end{itemize}

\subsection{Baseline Systems}

To evaluate performance, the proposed framework is compared against a cloud-only architecture:

\begin{itemize}
    \item \textbf{Cloud-Only Model:} All data generated by IoT devices is transmitted directly to the cloud for processing and for generation of the response returned to the device.

    \item \textbf{EdgeStream (Proposed):} Data is first processed at edge nodes, time-critical acknowledgement or actuation is produced locally, and only relevant information is forwarded to the cloud.
\end{itemize}

\subsection{Evaluation Procedure}

The experiments are conducted under varying workloads and system conditions:

\begin{itemize}
    \item Varying number of IoT devices (50--500)
    \item Increasing data rates (100--1000 msg/s)
    \item Comparison of edge vs. cloud performance under identical conditions
\end{itemize}

The workload generator spans the full stress envelope above, but the scenario tables in Section~\ref{sec:results} report stable operating points and saturation plateaus rather than the maximum configured generator rate. Specifically, the reported Healthcare, Industrial IoT, and Smart City profiles use 60, 70, and 75 devices, respectively, each evenly distributed across five logical processing nodes at 100\,msg/s/device. The corresponding mean serialized message sizes are 256\,B, 307\,B, and 378\,B, which remain within the stated 256\,B--1\,KB range. Higher-rate sweeps up to 1000\,msg/s/device are used only for the percentile-latency and aggregate-scaling stress tests discussed separately in the text.

Each experiment was repeated 10 times and average values are reported.

\subsection{Evaluation Metrics}

The system is evaluated using the following metrics:

\begin{itemize}
    \item \textbf{Latency:} Mean time from message generation to completion of the time-critical response, averaged over all messages in a run
    \item \textbf{Throughput:} Sustainable processing rate per active processing node, measured at the largest stable input rate before queue growth
    \item \textbf{Bandwidth Usage:} Average cloud-uplink traffic per device-hour at the nominal workload point
    \item \textbf{Energy Consumption:} Estimated based on device power usage
\end{itemize}

For the cloud-only baseline, the latency metric ends when the cloud response reaches the device. For EdgeStream, it ends when the edge node completes the local acknowledgement or control action; asynchronous summary uploads to the cloud are excluded from the latency-critical path.

Unless explicitly labelled as aggregate, the results in Section~\ref{sec:results} use these normalized definitions. Whole-system throughput can be recovered by multiplying the per-node value by the number of active processing nodes, and whole-system cloud traffic can be recovered by multiplying the per-device-hour value by the number of active devices.

The experimental results based on this setup are presented in Section~\ref{sec:results}.

\section{Results and Analysis}\label{sec:results}
Tables~\ref{tab:latency}, \ref{tab:throughput}, \ref{tab:stress-tests}, and \ref{tab:bandwidth}, together with Figs.~\ref{fig:latency-comparison} and \ref{fig:bandwidth-comparison}, summarise the evaluated performance of EdgeStream relative to the cloud-only baseline. Latency is reported per message, throughput per processing node, and bandwidth per device-hour; supplementary percentile-latency and raw-ingress stress tests are reported separately in Table~\ref{tab:stress-tests} with their workload basis. Percentage improvements are computed only within each metric's consistent reporting scope. Across the evaluated scenarios, EdgeStream reduced latency and cloud-uplink traffic while maintaining higher sustainable throughput.

\subsection{Latency Performance Analysis}
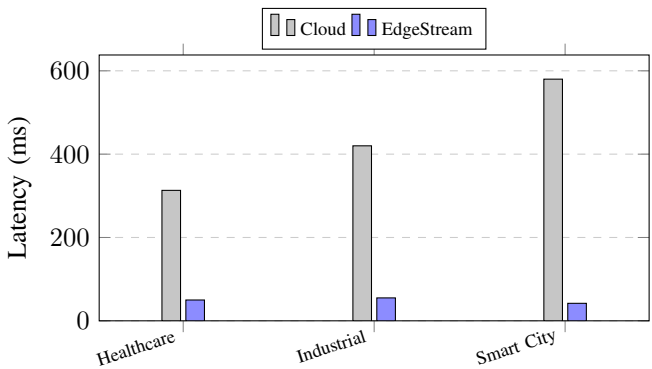
\begin{figure}[htbp]
    \centering
    \begin{tikzpicture}
        \begin{axis}[
            ybar,
            bar width=7pt,
            width=\columnwidth,
            height=5.1cm,
            ymin=0,
            ylabel={Latency (ms)},
            symbolic x coords={Healthcare,Industrial,SmartCity},
            xtick=data,
            xticklabels={Healthcare,Industrial,Smart City},
            xticklabel style={font=\scriptsize, rotate=18, anchor=east},
            ymajorgrids=true,
            grid style={dashed, gray!40},
            legend style={font=\scriptsize, at={(0.5,1.02)}, anchor=south, legend columns=2},
            enlarge x limits=0.22
        ]
            \addplot[fill=gray!45, draw=black] coordinates {
                (Healthcare,313.09)
                (Industrial,420.00)
                (SmartCity,580.00)
            };
            \addplot[fill=blue!45, draw=black] coordinates {
                (Healthcare,49.97)
                (Industrial,55.00)
                (SmartCity,42.00)
            };
            \legend{Cloud,EdgeStream}
        \end{axis}
    \end{tikzpicture}
    \caption{Latency comparison for the evaluated application scenarios.}
    \label{fig:latency-comparison}
\end{figure}

Table~\ref{tab:latency} and Fig.~\ref{fig:latency-comparison} show that edge-side processing lowers the mean end-to-end delay in all three application scenarios.

\begin{table}[htbp]
    \centering
    \caption{Latency comparison across application scenarios}
    \label{tab:latency}
    \footnotesize
    \setlength{\tabcolsep}{4pt}
    \begin{tabular}{lccc}
        \hline
        Scenario & Cloud (ms) & EdgeStream (ms) & Reduction (\%) \\
        \hline
        Healthcare & 313.09 & 49.97 & 84.0 \\
        Industrial IoT & 420.00 & 55.00 & 86.9 \\
        Smart City & 580.00 & 42.00 & 92.8 \\
        \hline
    \end{tabular}
\end{table}

\begin{itemize}
    \item In the healthcare scenario, average latency decreased from 313.09\,ms to 49.97\,ms.
    \item In the industrial IoT scenario, latency decreased from 420\,ms to 55\,ms.
    \item In the smart-city scenario, latency decreased from 580\,ms to 42\,ms, corresponding to the largest observed reduction.
    \item In the supplementary high-rate percentile sweep summarised in Table~\ref{tab:stress-tests}, EdgeStream remained below 50\,ms at the 95th percentile, whereas the cloud baseline had already exceeded 500\,ms by the 80th percentile.
\end{itemize}

The latency reduction is mainly associated with local processing at nearby edge nodes, which shortens the communication path and reduces network round-trip delay.

\subsection{Throughput and Scalability Performance}
Table~\ref{tab:throughput} summarises the sustainable per-node throughput results for the normalized application-message workloads. Table~\ref{tab:stress-tests} separately reports the supplementary raw-ingress and scaling stress tests together with their workload basis. Across the evaluated scenarios, EdgeStream sustained approximately twice as many application messages per second per processing node as the cloud-only baseline.

\begin{table}[htbp]
    \centering
    \caption{Sustainable throughput comparison across application scenarios (per active processing node)}
    \label{tab:throughput}
    \footnotesize
    \setlength{\tabcolsep}{3pt}
    \begin{tabular}{lccc}
        \hline
        Scenario & \shortstack{Cloud\\(msg/s/node)} & \shortstack{EdgeStream\\(msg/s/node)} & Gain ($\times$) \\
        \hline
        Healthcare & 1200 & 2400 & 2.00 \\
        Industrial IoT & 1400 & 2800 & 2.00 \\
        Smart City & 1500 & 3200 & 2.13 \\
        \hline
    \end{tabular}
\end{table}

\begin{table}[htbp]
    \centering
    \caption{Supplementary stress tests and workload basis}
    \label{tab:stress-tests}
    \scriptsize
    \setlength{\tabcolsep}{2pt}
    \begin{tabular}{p{1.45cm}p{2.75cm}p{3.55cm}}
        \hline
        Test & Workload basis & Reported observation \\
        \hline
        Percentile-latency sweep & Five logical processing nodes; synthetic streams; higher-rate sweep up to 1000\,msg/s/device as described in Section~\ref{sec:results} & EdgeStream 95th-percentile latency $< 50$\,ms; cloud baseline already $> 500$\,ms by the 80th percentile \\
        Single-node raw ingress & One optimised Raspberry Pi~4; synthetic raw sensor readings; stable point taken at the largest offered load with latency $< 30$\,ms & Approximately 15{,}000 readings/s \\
        Five-node scaling & Three physical edge devices plus two virtual instances; synthetic raw sensor readings; aggregate stable point before queue growth & Approximately 73{,}000 readings/s aggregate, with near-linear scaling over the evaluated range \\
        \hline
    \end{tabular}
\end{table}

\begin{itemize}
    \item In healthcare monitoring, sustainable per-node throughput increased from 1200\,msg/s/node to 2400\,msg/s/node.
    \item In the industrial IoT scenario, sustainable per-node throughput increased from 1400\,msg/s/node to 2800\,msg/s/node.
    \item In the smart-city scenario, sustainable per-node throughput increased from 1500\,msg/s/node to 3200\,msg/s/node.
    \item Table~\ref{tab:stress-tests} reports the supplementary raw-ingress stress tests separately from the normalized application-message throughput values in Table~\ref{tab:throughput}.
    \item In the single-node raw-ingress test, one optimised Raspberry Pi~4 sustained approximately 15{,}000 sensor readings/s at the largest offered load that kept latency below 30\,ms.
    \item In the five-node scaling test, the mixed physical/virtual configuration sustained approximately 73{,}000 raw sensor readings/s in aggregate over the evaluated range.
\end{itemize}

These results indicate that distributing computation across multiple edge nodes reduces the likelihood of the cloud becoming a bottleneck as the workload increases. For example, under the Healthcare nominal profile, the per-node cloud baseline of 1200\,msg/s corresponds to the offered load from 12 devices at 100\,msg/s/device, whereas EdgeStream provides roughly 2$\times$ headroom at the same operating point. The raw-ingress stress figures in Table~\ref{tab:stress-tests} use synthetic sensor readings and are therefore reported separately from the normalized application-message throughput values in Table~\ref{tab:throughput}.

\subsection{Bandwidth Utilisation and Savings}
\begin{figure}[htbp]
    \centering
    \begin{tikzpicture}
        \begin{axis}[
            ybar,
            bar width=7pt,
            width=\columnwidth,
            height=5.1cm,
            ymin=0,
            ylabel={Bandwidth (MB/device-hour)},
            symbolic x coords={Healthcare,Industrial,SmartCity},
            xtick=data,
            xticklabels={Healthcare,Industrial,Smart City},
            xticklabel style={font=\scriptsize, rotate=18, anchor=east},
            ymajorgrids=true,
            grid style={dashed, gray!40},
            legend style={font=\scriptsize, at={(0.5,1.02)}, anchor=south, legend columns=2},
            enlarge x limits=0.22
        ]
            \addplot[fill=gray!45, draw=black] coordinates {
                (Healthcare,92.4)
                (Industrial,110.4)
                (SmartCity,136.2)
            };
            \addplot[fill=green!45, draw=black] coordinates {
                (Healthcare,16.6)
                (Industrial,12.8)
                (SmartCity,17.7)
            };
            \legend{Cloud,EdgeStream}
        \end{axis}
    \end{tikzpicture}
    \caption{Cloud-uplink bandwidth comparison for the evaluated application scenarios, reported per device-hour.}
    \label{fig:bandwidth-comparison}
\end{figure}
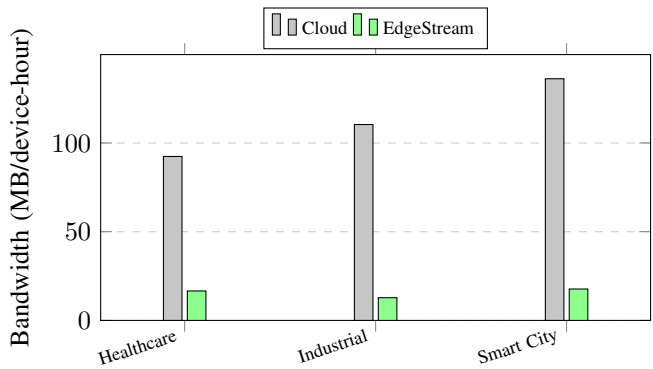

Table~\ref{tab:bandwidth} and Fig.~\ref{fig:bandwidth-comparison} show that edge-side filtering substantially reduces cloud-uplink traffic on a per-device-hour basis.

\begin{table}[htbp]
    \centering
    \caption{Cloud-uplink bandwidth comparison across application scenarios (per device-hour)}
    \label{tab:bandwidth}
    \footnotesize
    \setlength{\tabcolsep}{3pt}
    \begin{tabular}{lccc}
        \hline
        Scenario & \shortstack{Cloud\\(MB/device-h)} & \shortstack{EdgeStream\\(MB/device-h)} & Savings (\%) \\
        \hline
        Healthcare & 92.4 & 16.6 & 82.0 \\
        Industrial IoT & 110.4 & 12.8 & 88.4 \\
        Smart City & 136.2 & 17.7 & 87.0 \\
        \hline
    \end{tabular}
\end{table}

\begin{itemize}
    \item Cloud-uplink bandwidth decreased by 82--88\% across the evaluated scenarios.
    \item In the industrial IoT setting, EdgeStream used 12.8\,MB/device-hour compared with 110.4\,MB/device-hour for the cloud-only baseline.
    \item Under the Healthcare nominal profile (60 devices), the normalized values correspond to approximately 5.5\,GB/h of cloud-uplink traffic for the cloud-only baseline and about 1.0\,GB/h for EdgeStream.
\end{itemize}

Overall, the results indicate that EdgeStream is well suited to latency-sensitive and data-intensive IoT analytics, while still requiring a practical balance between edge and cloud resources.

\section{Discussion}

The experimental results indicate that the proposed \textit{EdgeStream} framework improves system performance relative to a traditional cloud-centric architecture under the evaluated conditions. The reduction in latency is primarily due to the proximity of edge nodes to IoT devices. By processing data locally, the system reduces communication delay and dependence on remote cloud infrastructure.

The increase in sustainable per-node throughput can be attributed to distributed processing across multiple edge nodes. Unlike centralized systems, where the cloud may become a bottleneck under high workloads, EdgeStream supports parallel data handling and scales more smoothly with increasing device counts.

Bandwidth savings are achieved by transmitting only filtered and relevant data to the cloud instead of raw data streams. When expressed as cloud-uplink traffic per device-hour, this selective communication reduces network load and makes the system more suitable for bandwidth-constrained environments, consistent with prior fog and edge-computing studies on communication reduction~\cite{shi2016edge,chiang2016fog,deng2016optimal}.

However, these improvements involve trade-offs. While edge processing reduces latency and bandwidth usage, it also introduces additional computational load on resource-constrained edge devices. This limits the complexity of algorithms that can be deployed locally. Therefore, a balance between edge and cloud processing remains important for strong overall performance~\cite{mach2017mec,you2017energy,deng2016optimal}.

\section{Limitations}

Despite promising results, the proposed framework has several limitations:

\begin{itemize}
    \item \textbf{Simulation-Based Evaluation:} The experiments are conducted in a simulated environment, which may not fully capture real-world challenges such as network instability, hardware failures, and environmental variability.

    \item \textbf{Limited Model Complexity:} The use of lightweight anomaly detection techniques restricts the ability to detect more complex patterns compared to advanced machine learning or deep learning models~\cite{chandola2009anomaly,li2018learning}.

    \item \textbf{Resource Constraints:} Edge devices have limited computational power, memory, and energy capacity, which may affect performance under high workloads or complex analytics tasks.

    \item \textbf{Scalability Validation:} Although the system demonstrates improved scalability in simulation, large-scale real-world deployment has not been fully validated.
\end{itemize}

These limitations highlight the need for further investigation and real-world testing to strengthen the applicability of the proposed approach.

\section{Conclusion}

This paper presented \textit{EdgeStream}, a low-latency edge-based framework for real-time IoT data analytics. The proposed system leverages a multi-layer architecture to process data closer to the source, reducing dependency on centralized cloud infrastructure.

Experimental results indicate that the framework can reduce mean end-to-end latency by up to 92.8\%, nearly double sustainable per-node throughput, and provide 82--88\% cloud-uplink bandwidth savings per device-hour under the evaluated conditions. These findings suggest that edge computing is a practical approach for scalable and efficient real-time analytics in IoT environments.

\section{Future Work}

Future research can extend this work in several directions:

\begin{itemize}
    \item \textbf{Advanced AI Integration:} Incorporating lightweight deep learning models and TinyML techniques to improve anomaly detection accuracy at the edge~\cite{li2018learning}.

    \item \textbf{Federated Learning:} Enabling collaborative model training across distributed edge nodes while preserving data privacy~\cite{mcmahan2017communication,atul2025federated}.

    \item \textbf{Real-World Deployment:} Validating the framework in large-scale, real-world IoT environments such as smart cities or industrial systems.

    \item \textbf{Adaptive Resource Management:} Developing intelligent mechanisms for dynamic task allocation between edge and cloud based on workload and network conditions, potentially incorporating agentic and self-managing cloud orchestration principles~\cite{sardellitti2015joint,you2017energy,deng2016optimal,shukla2026agentic}.

    \item \textbf{Security Enhancements:} Integrating advanced security mechanisms such as zero-trust architectures, secure communication protocols, and domain-specific intrusion-detection pipelines for critical infrastructures~\cite{roman2018security,MKDas2026}.
\end{itemize}

\end{document}